\documentclass[conference]{IEEEtran}
\IEEEoverridecommandlockouts
\usepackage{cite}
\usepackage{amsmath,amssymb,amsfonts}
\usepackage{algorithmic}
\usepackage{graphicx}
\usepackage{textcomp}
\usepackage{xcolor}
\usepackage{booktabs}
\usepackage{subcaption}
\usepackage{enumitem}
\usepackage[hidelinks]{hyperref}

\def\BibTeX{{\rm B\kern-.05em{\sc i\kern-.025em b}\kern-.08em
    T\kern-.1667em\lower.7ex\hbox{E}\kern-.125emX}}

\begin{document}

\title{Cross-Modal Semantic-Enhanced Diffusion Framework\\for Diabetic Retinopathy Grading}

\author{
    \IEEEauthorblockN{Yiqun Wang}
    \IEEEauthorblockA{
        \textit{School of Software Engineering} \\ 
        \textit{Beijing Jiaotong University} \\  
        Beijing, China \\
        23301076@bjtu.edu.cn  
    }
}
\maketitle

\begin{abstract}
\mdseries
Automated grading of diabetic retinopathy (DR) faces several critical
challenges: subtle inter-grade visual distinctions in fine-grained lesion
patterns, distributional discrepancies induced by heterogeneous imaging
devices and acquisition conditions, and the inherent inability of
purely visual approaches to exploit clinical semantic knowledge.
In this paper, we propose \textbf{CLIP-Guided Semantic Diffusion}
(\textbf{CGSD}), a DR grading framework that synergistically integrates
vision-language pretraining with diffusion probabilistic modeling.
We adopt a domain-specific vision-language model tailored for DR grading
as the semantic guidance module and adapt it to the target domain via
\textbf{Low-Rank Adaptation (LoRA)}, effectively bridging the distributional gap
between the pretrained model and the target dataset with only a minimal
number of trainable parameters.
Building on this foundation, we construct a \textbf{cross-modal semantic conditioning
vector} by computing the dot product between image features and the text
description features of each DR grade, yielding a joint representation that
simultaneously encodes visual content and clinical-grade semantics.
This vector serves as the conditioning signal for the diffusion denoising
network, replacing the structurally complex dual-branch visual prior
employed in existing diffusion-based classification methods.
Experiments on the APTOS 2019 dataset demonstrate that the proposed approach
achieves an \textbf{accuracy of 87.5\%} and a \textbf{macro-averaged F1 score of 0.731},
outperforming a variety of representative methods.
Ablation studies further validate the independent contribution of each
constituent module.
\end{abstract}

\begin{IEEEkeywords}
diabetic retinopathy, CLIP, diffusion model, LoRA
\end{IEEEkeywords}

\begin{figure*}[t]
    \centering
    \includegraphics[width=\textwidth]{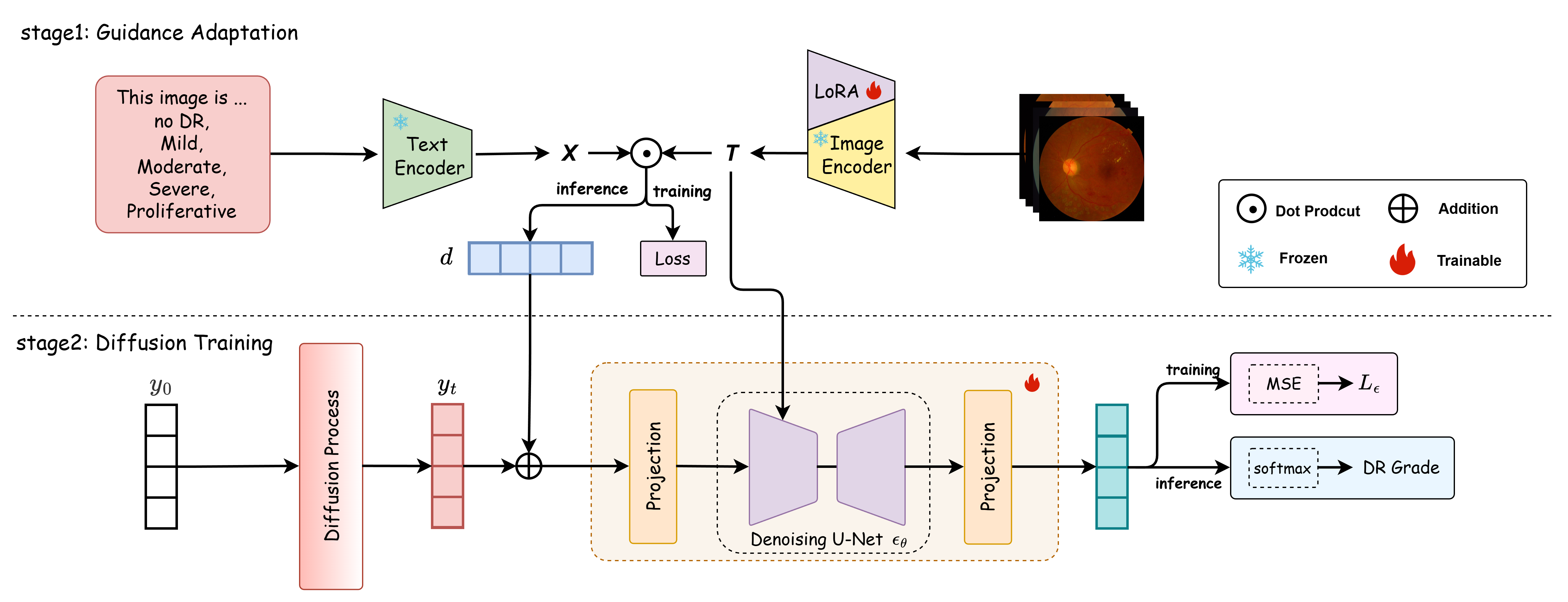}
    \caption{Overview of the proposed CGSD framework with two decoupled
    training stages.
    \textbf{Stage~1 (Guidance Adaptation):} A frozen Text Encoder and a
    LoRA-adapted Image Encoder extract $\ell_2$-normalized text features
    $\mathbf{T}$ and image features $\mathbf{f}$; their dot product yields
    the cross-modal semantic vector $\mathbf{d} \in \mathbb{R}^K$.
    After convergence, all guidance module parameters are frozen.
    \textbf{Stage~2 (Diffusion Training):} The ground-truth label $y_0$
    is progressively noised to $y_t$ and concatenated with $\mathbf{d}$
    as input to the trainable Denoising U-Net $\epsilon_\theta$, optimized
    by $\mathcal{L}_\epsilon$. During inference, starting from
    $y_T \sim \mathcal{N}(\hat{y}_0, \mathbf{I})$, the reverse denoising
    chain produces the final DR grade prediction via softmax.}
    \label{fig:model}
\end{figure*}

\section{Introduction}

Diabetic retinopathy (DR) is one of the leading causes of preventable
blindness worldwide.
With a continuously growing global patient population, early detection
and timely intervention are of critical importance for reducing
vision loss~\cite{teo2021global}.
Nevertheless, automated DR grading poses considerable practical
challenges~\cite{gulshan2016retinal}.
The disease manifests in diverse clinical forms, encompassing
microaneurysms, exudates, hemorrhages, and neovascularization, with
subtle visual differences between adjacent severity grades rendering
inter-grade discrimination particularly demanding.
Furthermore, fundus images acquired from different devices and imaging
environments exhibit pronounced variations in color style and contrast,
further compounding the difficulty of domain adaptation.

In recent years, deep learning methods have achieved remarkable advances
in automated DR grading.
Early efforts primarily relied on convolutional neural networks (CNNs) to
extract texture and spatial features from retinal
images~\cite{gulshan2016retinal}, attaining strong performance under
sufficiently annotated conditions.
DiffMIC~\cite{yang2023diffmic} subsequently extended diffusion probabilistic
models to medical image classification by employing a dual-granularity
conditional guidance strategy to model uncertainty in the label space,
validating the distinctive modeling capacity of diffusion denoising
mechanisms for fine-grained ordinal grading.
However, these approaches are fundamentally constrained by a purely
vision-driven supervised learning paradigm, precluding the exploitation
of semantic knowledge embedded in clinical textual descriptions.
This limitation is particularly pronounced in scenarios where the
pathological boundaries between adjacent grades rely heavily on semantic
conceptual distinctions for reliable differentiation.

Cross-modal approaches based on vision-language pretraining offer
a promising avenue to address these shortcomings.
CLIP-DR~\cite{yu2024clipdr} builds upon CLIP~\cite{radford2021clip} by
introducing a rank-aware prompting strategy that explicitly encodes the
natural ordinal relationships among DR grades into the image-text
alignment space, demonstrating superior grading discriminability across
multiple DR benchmarks.
Nevertheless, how to organically integrate such semantically
discriminative image-text features with the probabilistic denoising
capability of diffusion models remains insufficiently explored.
Motivated by this observation, we propose
\textbf{CLIP-Guided Semantic Diffusion} (\textbf{CGSD}), which adopts
CLIP-DR as a domain-specific guidance module, adapts it to the target
domain via LoRA, and constructs a cross-modal semantic conditioning
signal to replace the ROI-detection-dependent dual-branch visual prior
in DiffMIC~\cite{yang2023diffmic}.
This signal is directly injected into the diffusion denoising process,
achieving deep synergy between visual features and clinical semantics.

The main contributions of this paper are summarized as follows:
\begin{itemize}
    \item We propose a cross-modal semantic enhancement module that
    leverages a DR-oriented vision-language model to construct
    cross-modal conditioning signals in a joint semantic space,
    capturing both visual content and clinical-grade descriptions.
    This replaces the ROI-detection-dependent dual-branch visual prior
    in existing diffusion-based classification frameworks, simultaneously
    simplifying the overall architecture while enhancing semantic
    discriminability.

    \item We employ LoRA to adapt the domain-specific guidance model to
    the target domain, effectively bridging the distributional gap between
    the pretrained model and the APTOS target domain with only a small
    number of trainable parameters, while preserving the grade-discriminative
    capability acquired during pretraining.

    \item We design a two-stage decoupled training strategy that first
    ensures full convergence of the guidance module on the target dataset
    before independently training the diffusion classification network,
    thereby guaranteeing the stability of the conditioning signals.
    Systematic comparative experiments and ablation analyses on the
    APTOS 2019 dataset further validate the individual contribution
    of each module.
\end{itemize}

\section{Related Work}

\subsection{Vision-Language Pretraining for Medical Imaging}

Vision-language pretraining methods have demonstrated remarkable potential
in medical image analysis.
GLoRIA~\cite{huang2021gloria} proposed a global-local multimodal
representation learning framework that fully exploits paired radiology
images and diagnostic reports, enabling efficient recognition under
low-annotation conditions.
MedCLIP~\cite{wang2022medclip} extended contrastive learning to unpaired
medical image-text scenarios, substantially reducing dependence on
large-scale annotated data.
BiomedCLIP~\cite{zhang2023biomedclip} constructed a large-scale biomedical
foundation model pretrained on massive scientific image-text pairs,
demonstrating strong cross-modal generalization.
In ophthalmic image analysis, RET-CLIP~\cite{du2024retclip} pretrained a
dedicated retinal foundation model on fundus photographs paired with clinical
diagnostic reports, validating the efficacy of this paradigm for ocular
disease diagnosis including DR.
Building on this line of work, CLIP-DR~\cite{yu2024clipdr} further introduced
a rank-aware prompting mechanism that explicitly encodes the natural ordinal
relationships among DR grades into the image-text alignment space, achieving
superior fine-grained grading performance across multiple DR benchmarks.

\subsection{Diffusion Models for Medical Image Classification}

The application scope of diffusion probabilistic models has progressively
expanded from image generation to discriminative tasks such as medical
image classification.
DiffMIC~\cite{yang2023diffmic} pioneered the use of diffusion models in
medical image classification, modeling uncertainty in the label space through
a dual-granularity conditional guidance strategy and validating the
distinctive value of the diffusion denoising mechanism for fine-grained
medical grading.
However, these methods remain fundamentally constrained by a purely visual
supervised paradigm, making it difficult to incorporate clinical semantics
as high-level guidance signals.
This limitation motivates the present work to explore integrating
vision-language semantic knowledge into the diffusion classification framework.

\section{Methodology}

\subsection{Overall Framework}

As illustrated in Fig.~\ref{fig:model}, CGSD comprises two principal
components: a vision-language guidance module and a semantic-conditioned
diffusion classification network.
The guidance module is built upon CLIP-DR~\cite{yu2024clipdr}, which
augments CLIP with learnable Rank-aware Prompts and is jointly optimized
via a contrastive loss and a ranking loss,
$\mathcal{L}_{\text{total}} = \mathcal{L}_{\text{main}} +
\lambda\mathcal{L}_{\text{rank}}$, such that image and text features in the
shared semantic space satisfy the natural ordinal relationships among DR grades.
To address the distributional shift between the pretraining distribution of
CLIP-DR and the target domain APTOS~2019, we further apply
LoRA~\cite{hu2022lora} to adapt the guidance module.
With pretrained weights frozen, low-rank increment matrices
$\Delta W = BA$ ($r \ll \min(d,k)$) are injected into the high-semantic-level
layers, closing the domain gap with a minimal number of trainable parameters
while preserving the grade-discriminative capability acquired during
pretraining.
Upon convergence of the guidance module, it extracts the cross-modal
semantic conditioning signal $\mathbf{d}$, which is supplied to the
diffusion classification network for conditional denoising in the label
space to produce the final DR grade prediction.

To ensure that the diffusion network trains under stable conditioning
signals, we adopt a two-stage decoupled strategy.
In Stage~1, the guidance module is allowed to converge fully on the target
dataset, after which all its parameters are frozen.
In Stage~2, the frozen guidance module serves solely as a feature extractor,
and the denoising network $\epsilon_\theta$ is trained independently with
no gradient backpropagation to the guidance module.

\subsection{Cross-Modal Semantic Enhancement}
\label{subsec:enhancement}

Given an input image $x$, let the $\ell_2$-normalized image feature be
$\mathbf{f} \in \mathbb{R}^D$, and let the feature matrix of $K$ DR-grade
textual descriptions be
$\mathbf{T} = [f_t(p_1), \ldots, f_t(p_K)]^\top \in \mathbb{R}^{K \times D}$,
where $p_k$ denotes the Rank-aware Prompt for the $k$-th grade.
The cross-modal semantic feature vector is constructed via a dot product:
\begin{equation}
    \mathbf{d} = \mathbf{f} \cdot \mathbf{T}^\top \in \mathbb{R}^K
    \label{eq:d}
\end{equation}

The semantic discriminability of $\mathbf{d}$ stems from two complementary
sources.
First, the contrastive pretraining of CLIP-DR arranges image and text
features in the same embedding space according to semantic proximity;
consequently, $\mathbf{d}$ not only encodes the visual content of the image
but also explicitly captures the semantic association between the image and
each grade's linguistic description, incorporating cross-modal information
absent in purely visual features.
Second, the feature space refined by rank-aware fine-tuning satisfies
inter-grade ordinal constraints
($\mathcal{L}_{\text{total}} = \mathcal{L}_{\text{main}} +
\lambda\mathcal{L}_{\text{rank}}$), so that the individual components
of $\mathbf{d}$ naturally exhibit a monotonic trend consistent with the
pathological progression of DR, furnishing the diffusion classification
module with a semantically conditioned signal that possesses an intrinsic
ordinal structure.

In this work, $\mathbf{d}$ replaces the dual-branch visual prior that
relies on ROI detection in DiffMIC~\cite{yang2023diffmic}, directly
injecting visual-semantic alignment information into the diffusion process
while simplifying the overall architecture and providing the denoising
network with a more discriminative semantic conditioning signal.
Furthermore, $\hat{y}_0 = \text{softmax}(\mathbf{d})$ serves as the prior
mean of the diffusion forward process, permeating cross-modal semantic
knowledge into the noising procedure so that the reverse denoising commences
from a semantically informed initial distribution rather than uninformative
pure Gaussian noise.

\subsection{Diffusion-based Classification}
\label{subsec:diffusion}

Following the diffusion classification paradigm of
DiffMIC~\cite{yang2023diffmic}, we formulate DR grading as a
conditional denoising process in the label space.
The forward process takes $\hat{y}_0 = \text{softmax}(\mathbf{d})$ as the
prior mean and progressively adds noise to the ground-truth one-hot label
$y_0$ over $T$ steps:
\begin{multline}
    q(y_t \mid y_0, x)
    = \sqrt{\bar{\alpha}_t}\, y_0
      + \left(1 - \sqrt{\bar{\alpha}_t}\right)\hat{y}_0
      + \sqrt{1 - \bar{\alpha}_t}\,\boldsymbol{\epsilon}, \\
    \boldsymbol{\epsilon} \sim \mathcal{N}(\mathbf{0}, \mathbf{I})
    \label{eq:forward}
\end{multline}
where $\bar{\alpha}_t = \prod_{s=1}^{t}(1-\beta_s)$ and a linear noise
schedule is adopted.
Compared with DiffMIC, which conditions on
$(\rho(x), \hat{y}_g, \hat{y}_l, t)$, the proposed method replaces the
dual-branch visual prior with the semantic vector $\mathbf{d}$;
the conditioning input takes the form
$(\mathbf{f},\, y_t,\, \hat{y}_0,\, \mathbf{d},\, t)$,
and the training objective is:
\begin{equation}
    \mathcal{L}_{\text{diff}} =
    \mathbb{E}_{t,\, y_0,\, \boldsymbol{\epsilon}}
    \!\left[\,
    \left\|\boldsymbol{\epsilon} -
    \epsilon_\theta\!\left(\mathbf{f},\, y_t,\, \hat{y}_0,\,
    \mathbf{d},\, t\right)\right\|^2
    \right]
\end{equation}
During inference, a complete reverse denoising chain is executed starting
from $y_T \sim \mathcal{N}(\hat{y}_0, \mathbf{I})$; the mean of multiple
independent samples is subsequently passed through argmax to yield the
final DR grade prediction.

\section{Experiments}

\begin{table*}[t]
\centering
\caption{Quantitative Comparison with Baseline Methods on APTOS 2019.
The Best Results Are Marked in \textbf{Bold}.
Accuracy and Macro F1-Score Are Adopted as Metrics.}
\label{tab:sota}
\setlength{\tabcolsep}{8pt}
\begin{tabular}{l|cccccc|cc}
\hline\hline
\textbf{Methods}
  & LDAM~\cite{cao2019ldam} & OHEM~\cite{shrivastava2016ohem}
  & MTL~\cite{liao2018mtl}  & DANIL~\cite{gong2020danil}
  & CL~\cite{marrakchi2021cl} & ProCo~\cite{yang2022proco}
  & DiffMIC~\cite{yang2023diffmic} & \textbf{Ours} \\
\hline
Accuracy
  & 0.813 & 0.813 & 0.813 & 0.825 & 0.825 & 0.837
  & 0.858 & \textbf{0.875} \\
F1-Score
  & 0.620 & 0.631 & 0.632 & 0.660 & 0.652 & 0.674
  & 0.716 & \textbf{0.731} \\
\hline\hline
\end{tabular}
\end{table*}

\begin{table*}[t]
\centering
\caption{Ablation Study on APTOS 2019.
Each Row Adds One Component upon the Previous.
Best Results Are in \textbf{Bold}.}
\label{tab:ablation}
\setlength{\tabcolsep}{8pt}
\begin{tabular}{lcc}
\hline\hline
\textbf{Configuration} & \textbf{Accuracy (\%)} & \textbf{F1 Score} \\
\hline
Pretrained guidance model (no fine-tuning) & 77.3 & 0.540 \\
+ LoRA Fine-tuning                         & 84.7 & 0.686 \\
+ Semantic-Conditioned Diffusion           & \textbf{87.5} & \textbf{0.731} \\
\hline\hline
\end{tabular}
\end{table*}

\subsection{Dataset and Evaluation Metrics}

Experiments are conducted on the APTOS 2019 Diabetic Retinopathy Detection
dataset~\cite{karthik2019aptos}, which was collected at Aravind Eye Hospital,
India, and comprises 3,662 color fundus photographs.
Images are categorized into five severity grades: No DR, Mild, Moderate,
Severe, and Proliferative, exhibiting a pronounced class imbalance.
We strictly follow the DANIL protocol~\cite{gong2020danil} adopted by
DiffMIC~\cite{yang2023diffmic}, partitioning the data into training and test
sets at a 7:3 ratio to ensure a fair comparison with all baseline methods.

Two evaluation metrics are employed: accuracy (Acc), which measures the
overall fraction of correctly classified samples; and macro-averaged
F1 score (Macro F1), which computes the unweighted mean of per-class
F1 scores and thus effectively reflects balanced classification capability
under class-imbalanced conditions.

\begin{figure*}[t]
    \centering
    \includegraphics[scale=0.5]{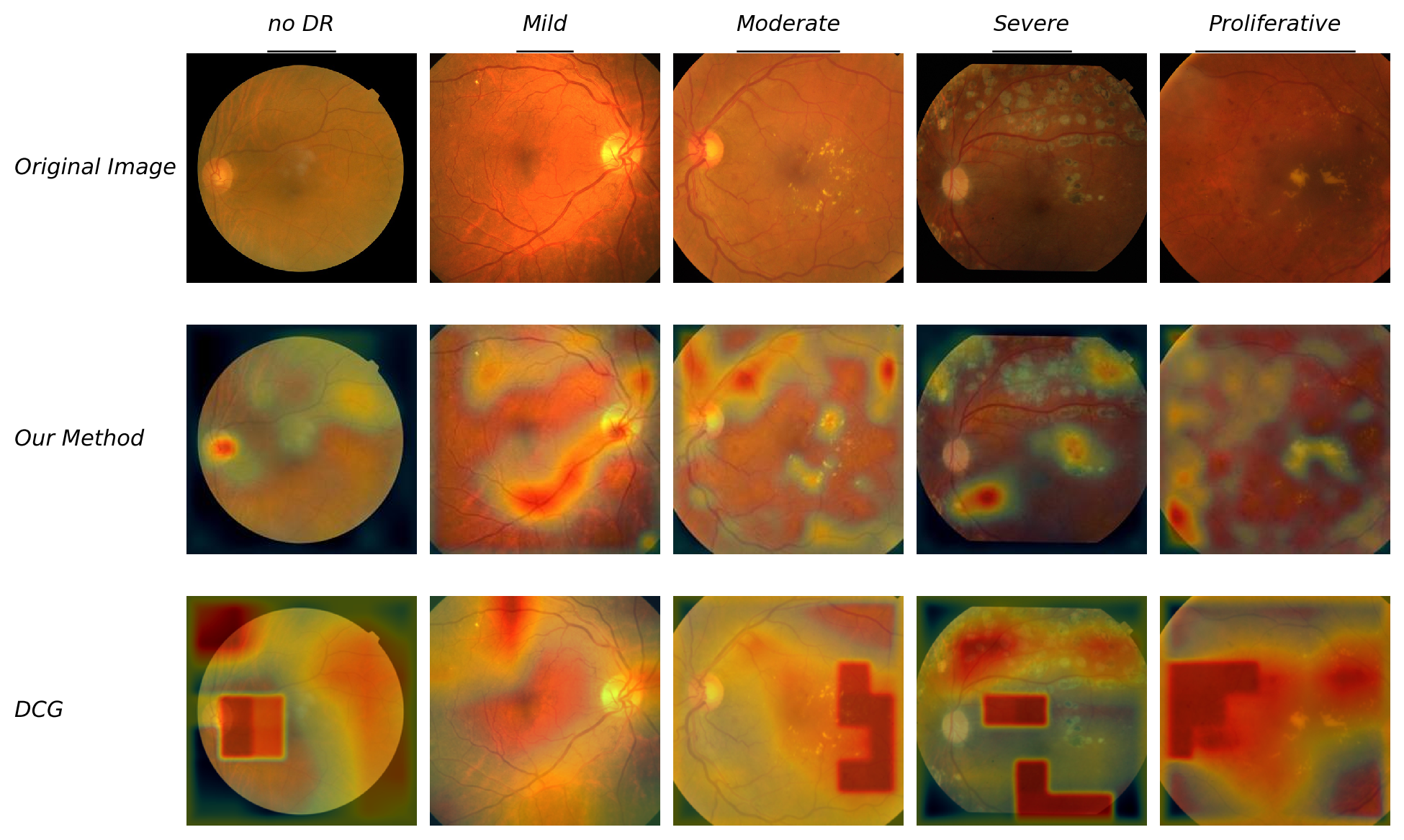}
    \caption{Class Activation Map (CAM) visualization on representative
    APTOS~2019 test samples across five DR severity grades.
    Row~1: original fundus images.
    Row~2: activation heatmaps from our LoRA fine-tuned guidance model
    (with cross-modal semantic conditioning).
    Row~3: activation heatmaps from the pretrained guidance model
    (without fine-tuning).
    The semantically conditioned model produces significantly more focused
    activations aligned with clinically relevant lesion regions,
    demonstrating that cross-modal semantic guidance steers the visual
    encoder toward pathology-aware feature extraction that pure visual
    supervision cannot achieve.}
    \label{fig:cam}
\end{figure*}

\begin{figure*}[t]
    \centering
    \includegraphics[width=\textwidth]{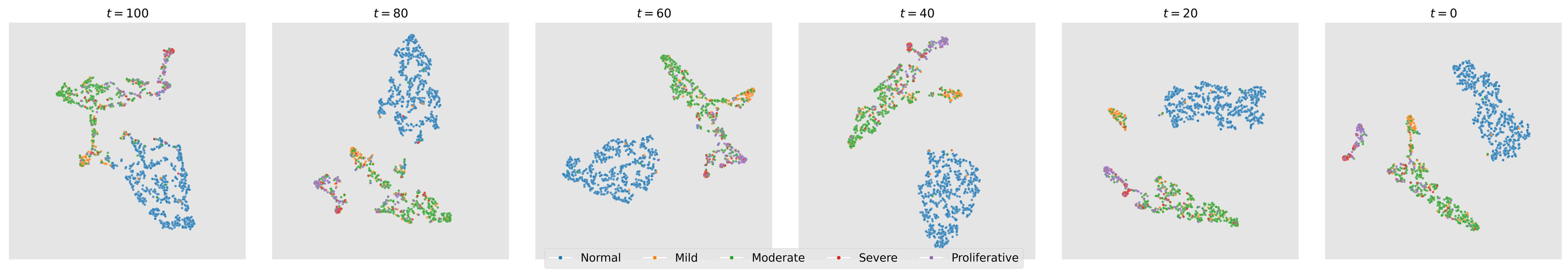}
    \caption{t-SNE visualization of label embeddings during the diffusion
    reverse process on the APTOS~2019 test set.
    As $t$ decreases from 100 to 0, class clusters become progressively
    more compact and separable, demonstrating that the
    semantic-conditioned denoising process effectively reduces prediction
    uncertainty across DR severity grades.}
    \label{fig:tsne_diffusion}
\end{figure*}

\subsection{Implementation Details}

\textbf{Guidance Module Fine-Tuning Stage.}
Starting from the CLIP-DR pretrained weights, LoRA fine-tuning is performed
on the APTOS~2019 training set.
LoRA is injected into the high-semantic-level layers of the visual encoder
(layer4 and Attention Pooling) with rank $r = 8$ and scaling factor
$\alpha = 16$.
The learning rate for LoRA increment parameters is set to $1\!\times\!10^{-4}$,
while Rank-aware Prompt parameters are optimized at $2\!\times\!10^{-3}$;
all remaining weights are kept strictly frozen.
The RAdam optimizer is employed with a cosine annealing learning rate
schedule, preceded by a 3-epoch linear warm-up starting from
$1\!\times\!10^{-5}$.
Training proceeds for 22 epochs with a batch size of 64;
input images are uniformly resized to $224\!\times\!224$ pixels, and
FP16 mixed-precision training is adopted.

\textbf{Diffusion Classification Network Training Stage.}
With the guidance module parameters frozen, a linear noise schedule is
applied ($T = 1000$, $\beta_1 = 1\!\times\!10^{-4}$ to
$\beta_T = 0.02$).
The denoising network $\epsilon_\theta$ is implemented as a fully
connected-layer-based conditional network.
The Adam optimizer ($\beta_1 = 0.9$) is used with an initial learning
rate of $3\!\times\!10^{-4}$, decayed to a minimum of
$1\!\times\!10^{-5}$, and gradient clipping at a threshold of 1.0
is applied.
Training proceeds for 500 epochs with a batch size of 32;
model weights are maintained via Exponential Moving Average (EMA) with
decay factor $\mu = 0.9999$.
During inference, five independent sampling runs are executed per image,
and the mean of the resulting predictions is passed through argmax to
obtain the final DR grade.
All experiments are performed on a single NVIDIA RTX~4090 GPU.

\subsection{Comparison with Baseline Methods}

To comprehensively evaluate the proposed method, CGSD is compared against
a set of representative baselines on the APTOS~2019 dataset, encompassing
loss-function-based long-tailed classification approaches
LDAM~\cite{cao2019ldam},
OHEM~\cite{shrivastava2016ohem}, and
MTL~\cite{liao2018mtl};
contrastive-learning-based methods
DANIL~\cite{gong2020danil},
CL~\cite{marrakchi2021cl}, and
ProCo~\cite{yang2022proco};
and the diffusion-model-based method DiffMIC~\cite{yang2023diffmic}.

As reported in Table~\ref{tab:sota}, purely classification-based methods
generally attain accuracy in the range of 81\%--84\%, while DiffMIC reaches
85.8\% by leveraging diffusion-model uncertainty modeling in the label space.
The proposed CGSD further advances performance to \textbf{87.5\%},
achieving the best results on both accuracy and F1 score.
Notably, CGSD attains a macro-averaged F1 of 0.730, a 1.4 percentage-point
improvement over DiffMIC's 0.716, indicating a substantive enhancement in
the recognition of minority-class grades under the highly imbalanced
five-class setting.
These results validate the efficacy of incorporating cross-modal semantic
conditioning signals into the diffusion classification framework.

\subsection{Ablation Study}

To assess the individual contribution of each module, we conduct a
stepwise ablation study on the APTOS~2019 test set; results are reported
in Table~\ref{tab:ablation}.
Starting from zero-shot inference using the pretrained guidance model
(Acc~77.3\%, F1~0.540), the introduction of LoRA target-domain adaptation
yields a substantial accuracy gain to 84.7\% (+7.4~pp), demonstrating that
even a small number of trainable parameters can effectively close the domain
gap and better align image features with the target-domain distribution.
Building upon this, the addition of the semantic-conditioned diffusion
classification framework further boosts accuracy to 87.5\% and F1 to 0.731,
contributing an additional 2.8 percentage points.
This confirms both the independent modeling value of the diffusion process
for predictive uncertainty and the effectiveness of the cross-modal semantic
conditioning signal for fine-grained ordinal grading.

\subsection{Visualization and Analysis}

\textbf{Class Activation Map Visualization.}
To analyze the influence of the cross-modal semantic conditioning signal
on the model's visual attention, we compare Class Activation Maps (CAMs)
before and after fine-tuning of the guidance model;
results are shown in Fig.~\ref{fig:cam}.
After LoRA fine-tuning and semantic-conditioned guidance, the model's
activation regions exhibit a considerably more concentrated and clinically
relevant spatial distribution.
In the Mild and Moderate grades, high-activation regions are precisely
focused on characteristic lesion sites such as exudates and microaneurysms;
in the Severe grade, the activation heatmap accurately covers the hemorrhage
and hard exudate areas at the center of the image;
in the Proliferative grade, high-activation regions are concentrated around
the optic disc, closely matching the typical spatial distribution of
neovascularization.
In contrast, the activation regions of the pretrained model without
fine-tuning are considerably more diffuse, lacking precise localization
of DR-specific lesions.

These differences originate from the deep influence of cross-modal
semantic conditioning on visual representation: the image-text alignment
training mutually constrains the grade descriptions on the language side
and image features within the same semantic space, guiding the visual
encoder to concentrate its attention on image regions semantically
consistent with the clinical characteristics of each DR grade.
This endows the model with pathology-aware feature extraction capability
that surpasses what purely visual supervision can provide.
Compared with the DCG module in DiffMIC, which depends on ROI detection
to obtain local priors, the semantic conditioning approach proposed in
this work requires no additional detection pipeline while offering stronger
semantic interpretability, maintaining stable lesion-perception ability even
when visual differences between adjacent grades are subtle.

\textbf{t-SNE Visualization of the Diffusion Reverse Process.}
Fig.~\ref{fig:tsne_diffusion} presents t-SNE visualizations of label
embeddings at different timesteps during the reverse denoising process.
As the denoising steps progress, inter-class structure gradually emerges:
grade clusters evolve from an unstructured noise distribution at $t=100$
to a discriminative layout with tighter boundaries and improved class
separability at $t=0$, providing direct visual evidence that the conditional
guidance of the cross-modal semantic feature $\mathbf{d}$ effectively
directs the diffusion model to resolve predictive uncertainty in the label
space.

%

\section{Conclusion}

This paper presents a joint classification framework for fine-grained DR
grading that integrates vision-language pretraining with diffusion
probabilistic modeling.
Existing methods are constrained by a purely visual supervised paradigm,
limiting the discriminability of fine-grained ordinal grading due to the
inability to incorporate clinical semantic knowledge.
To this end, we adopt a DR-oriented vision-language model as the semantic
guidance module, adapt it to the target domain via LoRA fine-tuning, and
construct a cross-modal semantic conditioning vector $\mathbf{d}$ that
explicitly injects the semantic alignment information between the input
image and each grade's clinical textual description into the diffusion
classification process, replacing the ROI-detection-dependent dual-branch
visual prior in the DiffMIC framework.
Experiments on the APTOS~2019 dataset demonstrate that the proposed method
achieves an accuracy of 87.5\%, surpassing a variety of representative
baseline methods, and ablation studies further validate the independent
contributions of both the LoRA fine-tuning stage and the
semantic-conditioned diffusion module.
These results indicate that combining vision-language alignment capability
with the probabilistic denoising mechanism of diffusion models constitutes
an effective pathway for advancing fine-grained medical image grading.

Future work will proceed along two directions.
First, the generalizability and transferability of the proposed framework
will be validated across additional DR datasets and other medical image
grading tasks.
Second, from the perspective of diffusion model optimization, we will
explore improvements to noise scheduling strategies and the introduction
of adaptive conditional guidance mechanisms to further enhance classification
accuracy and inference efficiency.



\begin{thebibliography}{00}

\bibitem{teo2021global}
Z. L. Teo, Y.-C. Tham, M. Yu, M. L. Chee, T. H. Rim, N. Cheung,
M. W. Bikbov, Y. X. Wang, Y. Tang, Y. Lu, I. Y. Wong, D. S. Ting,
G. S. Tan, J. B. Jonas, C.-Y. Cheng, and T. Y. Wong,
``Global prevalence of diabetic retinopathy and projection of burden
through 2045,''
\textit{Ophthalmology}, vol.~128, no.~11, pp.~1580--1591, 2021.

\bibitem{gulshan2016retinal}
V. Gulshan, L. Peng, M. Coram, M. C. Stumpe, D. Wu,
A. Narayanaswamy, S. Venugopalan, K. Widner, T. Madams,
J. Cuadros, R. Kim, R. Raman, P. C. Nelson, J. L. Mega,
and D. R. Webster,
``Development and validation of a deep learning algorithm for detection
of diabetic retinopathy in retinal fundus photographs,''
\textit{JAMA}, vol.~316, no.~22, pp.~2402--2410, 2016.

\bibitem{yang2023diffmic}
Y. Yang, H. Fu, A. I. Aviles-Rivero, C. Sch\"{o}nlieb, and L. Zhu,
``DiffMIC: Dual-guidance diffusion network for medical image
classification,''
in \textit{Proc. Int. Conf. Med. Image Comput. Comput.-Assist.
Interv.\ (MICCAI)}, Lecture Notes in Computer Science, vol.~14225,
2023, pp.~95--105.

\bibitem{yu2024clipdr}
Q. Yu, J. Xie, A. Nguyen, H. Zhao, J. Zhang, H. Fu, Y. Zhao,
Y. Zheng, and Y. Meng,
``CLIP-DR: Textual knowledge-guided diabetic retinopathy grading with
ranking-aware prompting,''
in \textit{Proc. Int. Conf. Med. Image Comput. Comput.-Assist.
Interv.\ (MICCAI)}, Lecture Notes in Computer Science, vol.~15009,
2024, pp.~662--671.

\bibitem{radford2021clip}
A. Radford, J. W. Kim, C. Hallacy, A. Ramesh, G. Goh, S. Agarwal,
G. Sastry, A. Askell, P. Mishkin, J. Clark, G. Krueger, and
I. Sutskever,
``Learning transferable visual models from natural language
supervision,''
in \textit{Proc. Int. Conf. Mach. Learn.\ (ICML)}, vol.~139, 2021,
pp.~8748--8763.

\bibitem{huang2021gloria}
S.-C. Huang, L. Shen, M. P. Lungren, and S. Yeung,
``GLoRIA: A multimodal global-local representation learning framework
for label-efficient medical image recognition,''
in \textit{Proc. IEEE/CVF Int. Conf. Comput. Vis.\ (ICCV)}, 2021,
pp.~3942--3951.

\bibitem{wang2022medclip}
Z. Wang, Z. Wu, D. Agarwal, and J. Sun,
``MedCLIP: Contrastive learning from unpaired medical images and
text,''
in \textit{Proc. Conf. Empir. Methods Nat. Lang. Process.\ (EMNLP)},
2022, pp.~3876--3887.

\bibitem{zhang2023biomedclip}
S. Zhang, Z. Xu, N. Usuyama, J. Bagga, R. Tinn, S. Preston, R. Rao,
M. Wei, N. Valluri, C. Wong, M. Lungren, T. Naumann, and H. Poon,
``BiomedCLIP: A multimodal biomedical foundation model pretrained from
fifteen million scientific image-text pairs,''
\textit{arXiv preprint arXiv:2303.00915}, 2023.

\bibitem{du2024retclip}
J. Du, J. Guo, Z. Ye, and M. Cheng,
``RET-CLIP: A retinal image foundation model pre-trained with clinical
diagnostic reports,''
\textit{arXiv preprint arXiv:2405.14137}, 2024.

\bibitem{hu2022lora}
E. J. Hu, Y. Shen, P. Wallis, Z. Allen-Zhu, Y. Li, S. Wang, L. Wang,
and W. Chen,
``LoRA: Low-rank adaptation of large language models,''
in \textit{Proc. Int. Conf. Learn. Represent.\ (ICLR)}, 2022.

\bibitem{cao2019ldam}
K. Cao, C. Wei, A. Gaidon, N. Arechiga, and T. Ma,
``Learning imbalanced datasets with label-distribution-aware margin
loss,''
in \textit{Proc. Adv. Neural Inf. Process. Syst.\ (NeurIPS)}, vol.~32,
2019.

\bibitem{shrivastava2016ohem}
A. Shrivastava, A. Gupta, and R. Girshick,
``Training region-based object detectors with online hard example
mining,''
in \textit{Proc. IEEE Conf. Comput. Vis. Pattern Recognit.\ (CVPR)},
2016, pp.~761--769.

\bibitem{liao2018mtl}
H. Liao and J. Luo,
``A deep multi-task learning approach to skin lesion classification,''
\textit{arXiv preprint arXiv:1812.03527}, 2018.

\bibitem{gong2020danil}
L. Gong, K. Ma, and Y. Zheng,
``Distractor-aware neuron intrinsic learning for generic 2D medical
image classifications,''
in \textit{Proc. Int. Conf. Med. Image Comput. Comput.-Assist.
Interv.\ (MICCAI)}, Lecture Notes in Computer Science, vol.~12261,
2020, pp.~591--601.

\bibitem{marrakchi2021cl}
Y. Marrakchi, O. Makansi, and T. Brox,
``Fighting class imbalance with contrastive learning,''
in \textit{Proc. Int. Conf. Med. Image Comput. Comput.-Assist.
Interv.\ (MICCAI)}, Lecture Notes in Computer Science, vol.~12903,
2021, pp.~466--476.

\bibitem{yang2022proco}
Z. Yang, J. Pan, Y. Yang, X. Shi, H.-Y. Zhou, Z. Zhang, and C. Bian,
``ProCo: Prototype-aware contrastive learning for long-tailed medical
image classification,''
in \textit{Proc. Int. Conf. Med. Image Comput. Comput.-Assist.
Interv.\ (MICCAI)}, Lecture Notes in Computer Science, vol.~13438,
2022, pp.~173--182.

\bibitem{karthik2019aptos}
S. D. Karthik and Maggie,
``APTOS 2019 blindness detection,''
Kaggle, 2019.
[Online]. Available:
\url{https://kaggle.com/competitions/aptos2019-blindness-detection}

\end{thebibliography}
\end{document}